\documentclass[twocolumn,showpacs,prb,floatfix]{revtex4}
\usepackage{epsfig}
\usepackage{tabularx}

\newcommand{\be}{\begin{equation}}
\newcommand{\ee}{\end{equation}}
\newcommand{\beqn}{\begin{eqnarray}}
\newcommand{\eeqn}{\end{eqnarray}}
\newcommand{\nsw}{N_{\mathrm{sweep}}}
\newcommand{\nsa}{N_{\mathrm{samp}}}

\newcommand{\chisg}{\chi_{_{\mathrm{SG}}}}

\begin{document}

\title{Monte Carlo studies of the one-dimensional Ising spin glass
with power-law interactions}
\author{Helmut G.~Katzgraber}
\altaffiliation{Present address:
Theoretische Physik, ETH H\"onggerberg, CH-8093 Z\"urich, Switzerland.}
\affiliation{Department of Physics, University of
California, Davis, California 95616}

\author{A.~P.~Young}
\email{peter@bartok.ucsc.edu}
\homepage{http://bartok.ucsc.edu/peter}
\affiliation{Department of Physics,
University of California,
Santa Cruz, California 95064}

\date{\today}

\begin{abstract}
We present results from Monte Carlo simulations of the one-dimensional
Ising spin glass with power-law interactions at low temperature, using the
parallel tempering Monte Carlo method. For a set of parameters where the
long-range part of the interaction is relevant, we find evidence for
large-scale dropletlike excitations with an energy that is independent of
system size, consistent with replica symmetry breaking. We also perform
zero-temperature defect energy calculations for a range of parameters and
find a stiffness exponent for domain walls in reasonable but by no means
perfect agreement with analytic predictions.
\end{abstract}

\pacs{75.50.Lk, 75.40.Mg, 05.50.+q}
\maketitle

\section{Introduction}
\label{introduction}

Two main theories of the spin-glass state have been proposed: the
``droplet picture,''\cite{fisher:86,fisher:87,fisher:88,bray:86} and the
replica symmetry-breaking (RSB) 
picture.\cite{parisi:79,parisi:80,parisi:83,mezard:87} The droplet picture
states that the minimum-energy excitation of linear dimension $l$
containing a given spin (a droplet) has an energy proportional to
$l^{\theta}$, where $\theta$ is positive (assuming that the transition
temperature $T_c$ is finite). It follows that in the thermodynamic limit,
excitations that flip a finite fraction of the spins cost an infinite
energy. In addition, the droplet picture states that these excitations are
fractal with a fractal dimension $d_s < d$, where $d$ is the space
dimension. By contrast, RSB follows Parisi's exact solution of the
Sherrington-Kirkpatrick (SK) infinite-range model and predicts that
excitations involving a finite fraction of the spins cost a finite energy
in the thermodynamic limit, and are space filling,\cite{marinari:00} i.e.,
$d_s = d$.

There have been several numerical
attempts\cite{krzakala:00,palassini:00,marinari:00,katzgraber:01,katzgraber:01a,katzgraber:01c,houdayer:00,houdayer:00b}
to better understand the nature of the spin-glass state for short-range
spin glasses. The data, which are on small system sizes, are consistent
with a picture in which the ``stiffness exponent'' for dropletlike
excitations is different from that for domain-wall excitations, 
although it is still
unclear if the difference is due to the limited range of system
sizes,\cite{moore:02} or if it persists in the thermodynamic limit.
Recently Krzakala and Martin,\cite{krzakala:00} as well as Palassini and
Young,\cite{palassini:00} find evidence for an intermediate picture
(called ``TNT'' which stands for trivial-nontrivial) in which only a
finite amount of energy is needed to generate system-size excitations in
the thermodynamic limit, but their surface has a nontrivial fractal
dimension less than $d$.

In addition to finite-$T$ Monte Carlo simulations, there has been a
substantial amount of numerical work studying the energy of domain walls
at
$T=0$.\cite{bray:84,mcmillan:84,mcmillan:84b,rieger:96,hartmann:99,palassini:99,hartmann:01a,carter:02}
The energy of the domain wall is found to vary as $L^\theta$ where $L$ is
the system size and $\theta$ is positive (for systems with $T_c > 0$). 
The droplet theory assumes that the stiffness exponent for domain walls
$\theta_{\rm domain-wall}$ is the same as the stiffness exponent for
dropletlike excitations $\theta_{\rm droplet}$. In addition, 
$\theta_{\rm domain-wall} > 0$. This is in contrast to the RSB and TNT
pictures where the exponents are expected to be different and 
$\theta_{\rm droplet} = 0$. To simplify the notation,\cite{comment:1} 
from now on we will denote
$\theta_{\rm domain-wall}$ by $\theta$ and $\theta_{\rm droplet}$ by
$\theta'$.

In this work we consider the question of whether there are one or two
stiffness exponents for the case of Ising spin-glass models in
one-dimension in which the interactions fall off with a power $\sigma$
of the distance. Depending on $\sigma$, the system can have either a
finite transition temperature $T_c$ (with the long-range part of the
interaction relevant), or $T_c = 0$, in which case it can be either in the
short-range or long-range universality class.\cite{bray:86b,fisher:88}
The advantages of this
system are the following: 
(i) a large range of system sizes $L$ can be studied, (ii)
there are analytical predictions for the stiffness exponent within the
droplet picture, and (iii) within a single model, simply by changing a
parameter, one can cover the full range of behavior, from the absence of a
spin-glass phase to the infinite-range SK model. We perform both Monte
Carlo simulations at low $T$ (which indirectly gives $\theta'$), for a
value of $\sigma$ that has a finite $T_c$, and ground-state calculations
for a range of values of $\sigma$ corresponding to all three regimes: $T_c
> 0,\ T_c = 0$ (long range), and $T_c=0$ (short range). We use the
parallel tempering\cite{hukushima:96,marinari:98b} approach for both the
finite-$T$ and $T=0$ studies.

In Secs.~\ref{model}, \ref{observables}, and \ref{equilibration}, we
introduce the model, observables, and details of the Monte Carlo
technique, respectively. In addition we describe the phase diagram of the
model in detail. The critical region of the disordered long-range Ising
spin glass is analyzed in Sec.~\ref{results1} in order to determine the
critical temperature of the system and to ensure that in
Sec.~\ref{results2}, where we study the system at finite $T$, the
temperature is sufficiently far below $T_c$ such that critical-point
fluctuations do not affect the data. The zero-temperature results are
described in Sec.~\ref{resultsT0} and our conclusions are summarized in
Sec.~\ref{conclusions}.

\section{Model}
\label{model}

\begin{figure}
\centerline{\epsfxsize=6cm \epsfbox{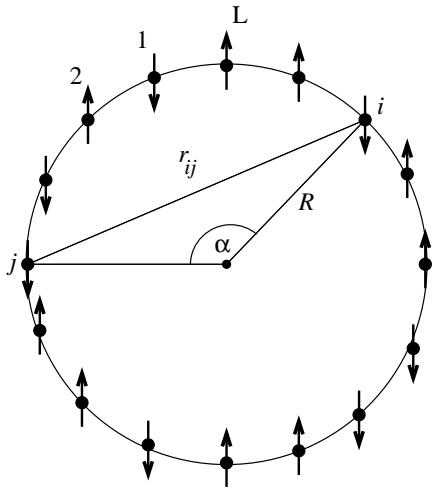}}
\caption{
Geometry of the one-dimensional Ising chain: the Ising spins are placed
equidistantly on a ring in order to ensure periodic boundary conditions.  
Their distance $r_{ij}$ is determined by calculating the geometric
distance between the different sites of the circle of radius $R$ and
circumference $L$ (not all bonds are shown for clarity).
}
\label{chain}
\end{figure}

The Hamiltonian for the one-dimensional long-range Ising spin glass 
with power-law interactions is given by
\begin{equation}
{\cal H} = -\sum_{i,j} J_{ij} S_i S_j \; ,
\label{hamiltonian}
\end{equation}
where the sites $i$ lie on a ring of length $L$, as shown in
Fig.~\ref{chain}, and $S_i = \pm 1$ represent Ising spins.  We place the
spins on a ring in order to ensure periodic boundary conditions. The sum
is over all spins on the chain and the couplings $J_{ij}$ are given by
\begin{equation}
J_{ij} = c(\sigma)\frac{\epsilon_{ij}}{r_{ij}^\sigma}\; ,
\label{bonds}
\end{equation}
where the $\epsilon_{ij}$ are chosen according to a Gaussian distribution
with zero mean and standard deviation unity:
\begin{equation}
{\mathcal P}(\epsilon_{ij}) = \frac{1}{\sqrt{2\pi}}\exp(-\epsilon_{ij}^2/2) \;.
\label{gaussian}
\end{equation}
The constant $c(\sigma)$ in Eq.~(\ref{bonds}) is chosen to give a mean-field
transition temperature $T_c^{\rm MF} = 1$, where
\begin{equation}
\left(T_c^{\rm MF}\right)^2 = \sum_{j\ne i} [ J_{ij}^2]_{\rm av} = 
c(\sigma)^2 \sum_{j\ne i} {1 \over r_{ij}^{2\sigma}} \; . 
\label{tcmf}
\end{equation}
Here $[\cdots]_{\rm av}$ denotes an average over disorder.

\begin{figure}
\centerline{\epsfxsize=7cm \epsfbox{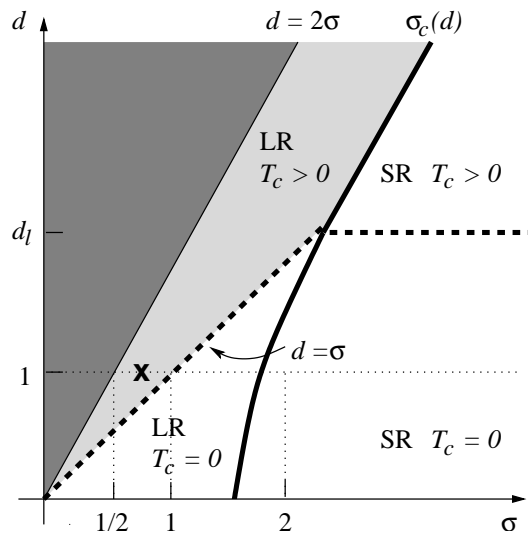}}
\caption{
Sketch of the phase diagram in the $d$-$\sigma$ plane for the spin-glass
state of the disordered long-range Ising model with power-law interactions
following Ref.~\onlinecite{fisher:88}. The light shaded region is where
there is both a finite $T_c$ and the spin-glass state is controlled by the
long-range part of the interaction (i.e., $\theta_{\rm LR} > \theta_{\rm
SR}$ and $\theta_{\rm LR} > 0$). The thick solid line separates the region
of short-range behavior from that of long-range behavior, i.e., it
corresponds to $\theta_{\rm SR} = \theta_{\rm LR}$, and is denoted by
$\sigma_c(d)$, see Eq.~(\ref{sigmac}). The thick dashed line separates
regions where $T_c = 0$ from regions where $T_c > 0$, i.e., it corresponds
to $\theta = 0$, where $\theta$ refers to the greater of $\theta_{\rm SR}$
and $\theta_{\rm LR}$. The dark shaded region is where there is no
thermodynamic limit unless the interactions are scaled appropriately by
system size. The calculations are performed for $d=1$ (marked by a
horizontal dotted line), for which (i) $\sigma_c(d) = 2$, (ii) $\theta >
0$ for $\sigma < 1$, and (iii) the thermodynamic limit does not exist for
$\sigma < 1/2$ (unless the interactions are scaled by a power of the
system size). These values of $\sigma$ are marked. The simulations at
finite $T$ are performed for $\sigma=3/4, d=1$, which is indicated by a
cross.
}
\label{dsigma}
\end{figure}

The distance between two spins on the chain $r_{ij}$ is determined by
$r_{ij} = 2R\sin(\alpha/2)$, where $R$ is the radius of the chain and
$\alpha$ is the angle between the two sites on the circle (see
Fig.~\ref{chain}).  The previous expression can be rewritten in terms of
$L$ and the positions of the spins to obtain
\begin{equation}
r_{ij} = \frac{L}{\pi}\sin\left(\frac{\pi |i - j|}{L}\right)\; .
\end{equation}

The long-range Ising spin glass with power-law interactions has a very
rich phase diagram in the $d$-$\sigma$ plane. This is summarized in
Fig.~\ref{dsigma}, which is based on the work by Bray {\em et al}.~\cite{bray:86b}
and by Fisher and Huse\cite{fisher:88} who present a detailed analysis of
the role of long-range interactions within the droplet model. Spin-glass
behavior is controlled by the long-range part of the interaction if
$\sigma$ is sufficiently small, and by the short-range part if $\sigma$ is
sufficiently large. More precisely, one has long-range behavior if the
stiffness exponent of the long-range (LR) universality class, $\theta_{\rm
LR}$, is greater than that of the short-range (SR) universality class,
$\theta_{\rm SR}$, and vice versa. In addition, there is an exact result
for $\theta_{\rm LR}$, namely,\cite{bray:86b,fisher:88}
\begin{equation}
\theta_{\rm LR} = d - \sigma ,
\label{theta_lr}
\end{equation}
so long-range behavior occurs if
\begin{equation}
\sigma < \sigma_c(d) = d - \theta_{\rm SR}(d)\; .
\label{sigmac}
\end{equation}
Equation (\ref{theta_lr}) indicates that critical exponents depend
continuously on $\sigma$ in the long-range region, even though they are
independent of $\sigma$ in the region controlled by the short-range part
of the interaction.

\begin{figure}
\centerline{\epsfxsize=7cm \epsfbox{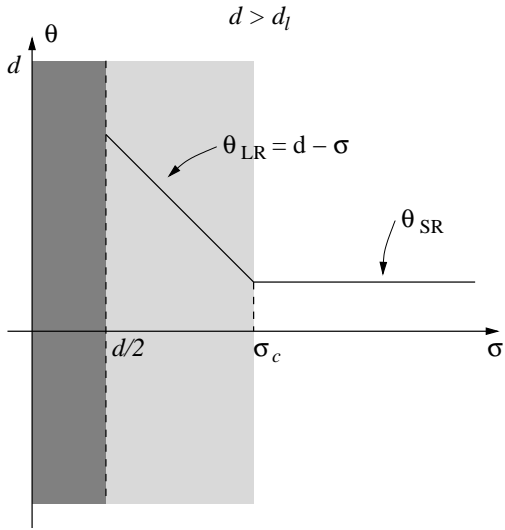}}
\caption{
Stiffness exponent $\theta$ as a function of $\sigma$ for $d > d_l$, where
$d_l$ is the lower critical dimension for short-range interactions. The
dark shaded region corresponds to $\sigma < d/2$ where $T_c$ diverges
unless the interactions are scaled by a power of the system size. The
light shaded region is where both $T_c > 0$ and the behavior is governed
by the long-range part of the interaction.
}
\label{dgtdl}
\end{figure}

\begin{figure}
\centerline{\epsfxsize=7cm \epsfbox{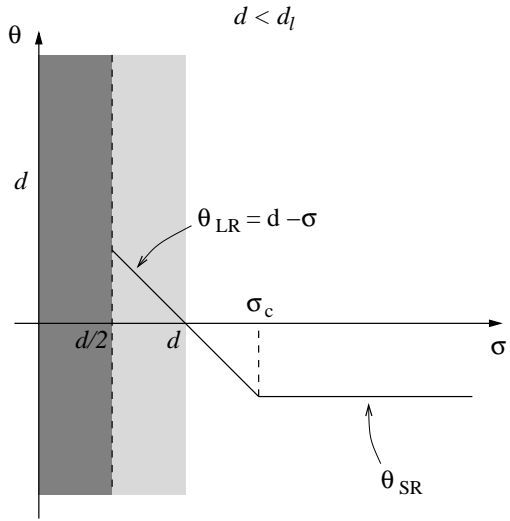}}
\caption{
Same as Fig.~\ref{dgtdl} but for $d < d_l$.
}
\label{dltdl}
\end{figure}

The condition for a finite-temperature transition is $\theta > 0$, where
$\theta$ refers here to the greater of $\theta_{\rm SR}$ and $\theta_{\rm
LR}$. For the short-range model, there is a finite-temperature transition
(i.e., $\theta_{\rm SR} > 0$) for $d$ larger than the lower critical
dimension $d_l$, which is found numerically to lie between 2 and
3.\cite{bray:84,mcmillan:84,mcmillan:84b,kawashima:96,ballesteros:00}
In Figs.~\ref{dgtdl} and \ref{dltdl}, we show the expected variation of
$\theta$ with $\sigma$ both for $d > d_l$ and $d < d_l$.

For $\sigma < d/2$ the model does not have a thermodynamic limit ($T_c$
diverges) unless the interactions are scaled with an appropriate (inverse)
power of $L$, i.e., $c(\sigma) \to 0 $ for $L \to \infty$. Note that
$\sigma = 0$ corresponds to the SK model.

For $d = 1$, the case studied in our numerics, one has\cite{bray:84}
\begin{equation}
\theta_{\rm SR}(1) = -1,
\label{theta_SR}
\end{equation}
and so from Eq.~(\ref{sigmac}), $\sigma_c = 2$, in agreement with rigorous
results by van Enter and van Hemmen.\cite{enter:85} This situation
corresponds to Fig.~\ref{dltdl} with $\theta_{\rm SR} = -1$. To obtain a
finite $T_c$ in $d=1$, one needs $1/2 < \sigma < 1$ and, in the finite-$T$
simulations, we choose $\sigma = 0.75$ (see Fig.~\ref{dsigma}).
For the zero-temperature calculations
we use a range of $\sigma$ between $0.10$ and $3.00$. From
Figs.~\ref{dsigma} and \ref{dltdl}, we see that this includes all the
possible types of behavior: (i) $T_c > 0$, (ii) $T_c = 0$ (LR), (iii) $T_c
= 0$ (SR), and (iv) SK-like behavior for $\sigma \rightarrow 0$.

\section{Observables}
\label{observables}

In our finite-$T$
simulations we primarily focus our attention on the spin overlap
defined by
\begin{equation}
q = {1 \over L} \sum_{i=1}^L S_i^{\alpha} S_i^{\beta} ,
\label{q}
\end{equation}
where ``$\alpha$'' and ``$\beta$'' refer to two copies (replicas) of the
system with the same disorder. In order to determine the critical
temperature of the system, we study the Binder ratio\cite{binder:81} $g$
defined by
\begin{equation}
g = \frac{1}{2}\left(3 - 
\frac{[\langle q^4\rangle]_{\rm av}}{[\langle q^2\rangle]_{\rm av}^2}\right)
= \tilde{g}[ L^{1/\nu}(T - T_c)] \; .
\label{g}
\end{equation}
Here, $\langle \cdots \rangle$ represents a thermal average, $\nu$ is the
correlation length exponent, and $T_c$ is the critical temperature of the
system. Because $g$ is dimensionless, curves for $g(T,L)$ cross at the
same $T_c$ for all $L$. We also compute the spin-glass susceptibility
$\chisg$, where
\begin{equation}
\chisg = L[\langle q^2 \rangle]_{\rm av}
= L^{2- \eta} \tilde{\chi}[L^{1/\nu}(T - T_c)] \; ,
\label{chi}
\end{equation}
and $\eta$ describes the power-law falloff of the spin-glass correlations
at $T_c$.

\section{Equilibration}
\label{equilibration}

\begin{figure}
\centerline{\epsfxsize=\columnwidth \epsfbox{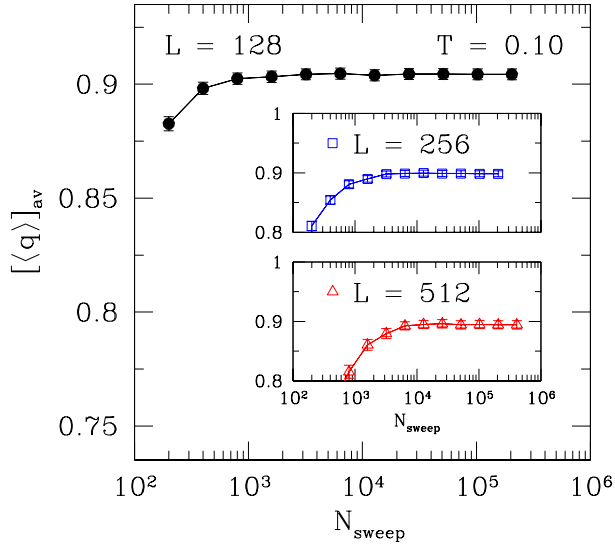}}
\vspace{-1.0cm}
\caption{
Average spin overlap $[\langle q \rangle]_{\rm av}$ as a function of Monte
Carlo sweeps $\nsw$, which each of the replicas perform, averaged over the
last half of the sweeps for different system sizes $L$. Note that the data
appear to be independent of the number of sweeps. The data shown are for $T
= 0.10$, the lowest temperature studied.
}
\label{equil}
\end{figure}

For the finite-$T$ simulations, we use the parallel tempering Monte Carlo
method\cite{hukushima:96,marinari:98b} as it allows us to study larger
systems at lower temperatures than is possible with conventional 
``single-spin-flip'' Monte Carlo methods. We test equilibration by the traditional
technique of requiring that different observables are independent of the
number of Monte Carlo steps $\nsw$.  Figure \ref{equil} shows data for the
spin overlap as a function of Monte Carlo steps for different system
sizes. We equilibrate by repeatedly doubling the number of Monte Carlo
sweeps until the last three agree within error bars.

We also checked the equilibration of some of our data using a
generalization of a test that was developed earlier\cite{katzgraber:01}
for nearest-neighbor models. For a Gaussian distribution of bonds it is
straightforward to show, by integrating by parts with respect to the
bonds, that the energy per spin, $U = -N^{-1} \sum_{\langle i, j\rangle} [
J_{ij} \langle S_i S_j \rangle ]_{\rm av},$ can be related to a
generalized ``link overlap'' $q_l$, defined by
\begin{equation}
q_l = {2 \over N}\sum_{\langle i,j\rangle } {[J_{ij}^2]_{\rm av} \over
(T_c^{MF})^2} 
[ \langle S_i S_j \rangle^2 ]_{\rm av} ,
\label{ql}
\end{equation}
as follows:
\begin{equation}
q_l = 1 - {2 T |U| \over (T_c^{\rm MF})^2}.
\label{equil_test}
\end{equation}
Note that this definition of $q_l$ is the natural generalization, to
long-range interactions, of the link overlap discussed in earlier
work\cite{palassini:00,katzgraber:01} on nearest-neighbor models, and that
$q_l$ is normalized so that it tends to unity for $T \to 0$ where $\langle
S_i S_j \rangle^2 \to 1$.

\begin{figure}
\centerline{\epsfxsize=\columnwidth \epsfbox{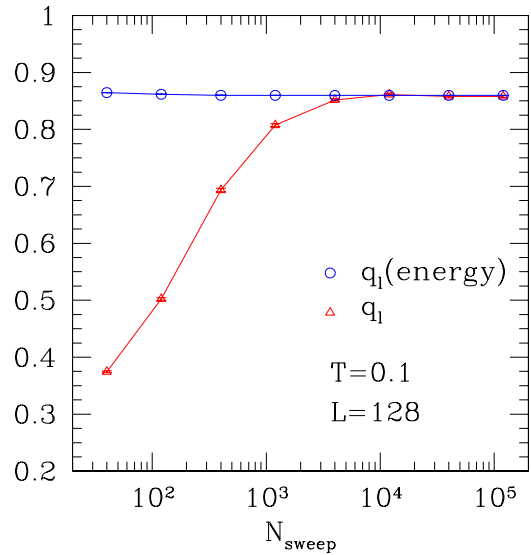}}
\caption{
Test for equilibration using Eq.~(\ref{equil_test}) as discussed in
Ref.~\onlinecite{katzgraber:01} and the text. The data are for $L=128,
T=0.1$. The figure shows both the link overlap $q_l$ defined in
Eq.~(\ref{ql}) and $q_l{\rm(energy)}$ which is the RHS of
Eq.~(\ref{equil_test}), for an increasing number of sweeps $N_{\rm sweep}$
that each replica performs.
}
\label{qltest}
\end{figure}

As discussed in Ref.~\onlinecite{katzgraber:01}, the two sides of
Eq.~(\ref{equil_test}) are only equal to each other in equilibrium and
approach the equilibrium value from opposite sides. This is illustrated in
Fig.~\ref{qltest} which shows $q_l$ and the quantity on the right-hand side
(RHS) of
Eq.~(\ref{equil_test}) for an increasing number of sweeps for $L=128,
T=0.1$. The two quantities do indeed approach each other from opposite
sides, and once they agree they do not change if the simulation is run for
longer. We applied this test for several lattice sizes using the
parameters in Table~\ref{simparams}, and found that the data were well
equilibrated in each case.

We also require that the acceptance ratios for the global moves which
interchange the different temperatures in the parallel tempering scheme be
greater than $0.4$ on average and roughly constant as a function of
temperature. Table \ref{simparams} shows the number of samples $\nsa$ and
the number of Monte Carlo sweeps $\nsw$ performed by each replica for each
lattice size. The lowest temperature used is $T = 0.10$, the highest $T =
1.40$, well above the mean-field critical temperature $T_c^{\rm MF} = 1$.

\begin{table}
\caption{
Parameters of the finite-$T$ simulations. $\nsa$ is the number of samples,
i.e., sets of disorder realizations, $\nsw$ is the total number of sweeps
simulated for each of the $2 N_T$ replicas for a single sample, and $N_T$
is the number of temperatures used in the parallel tempering method.
\label{simparams}
}
\begin{tabular*}{\columnwidth}{@{\extracolsep{\fill}} c r r l }
\hline
\hline
$L$  &  $\nsa$  & $\nsw$ & $N_T$  \\ 
\hline
32  & $2.0 \times 10^4$ & $1.0 \times 10^4$ & 23 \\
64  & $2.0 \times 10^4$ & $1.0 \times 10^4$ & 23 \\
128 & $2.0 \times 10^4$ & $2.0 \times 10^4$ & 23 \\
256 & $1.0 \times 10^4$ & $1.0 \times 10^5$ & 23 \\
512 & $5.0 \times 10^3$ & $2.0 \times 10^5$ & 23 \\
\hline
\hline
\end{tabular*}
\end{table}

We also use parallel tempering as an optimization algorithm to determine
the ground state in the $T=0$ calculations. This is described in
Sec.~\ref{resultsT0}.

\section{Finite-Size Scaling in the Critical Region}

\begin{figure}
\centerline{\epsfxsize=\columnwidth \epsfbox{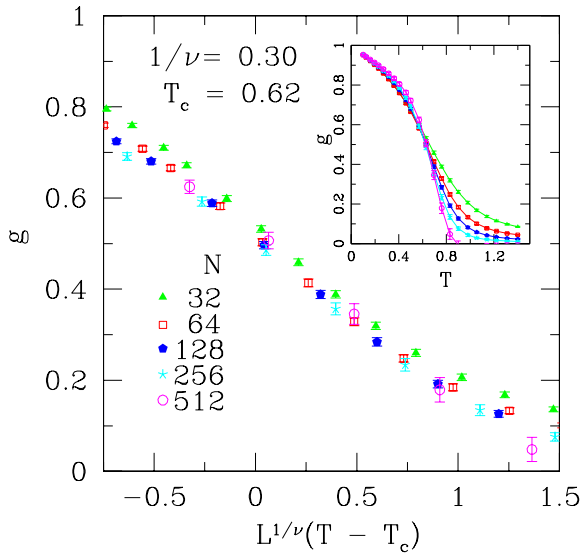}}
\vspace{-1.0cm}
\caption{
The Binder ratio $g$ as a function of $L^{1/\nu}(T - T_c)$ for several
system sizes. We see that the data scale well for $1/\nu = 0.30 \pm 0.03$
and $T_c = 0.62 \pm 0.03$. This is a little lower than the (approximate)
estimate of Bhatt and Young (Ref.~\onlinecite{bhatt:86})
who find $T_c \approx 0.68$.
The inset shows the unscaled data for $g$. One can see that the data for
different sizes cross at $T \sim 0.62$.
}
\label{binder}
\end{figure}

In this section, we determine the critical temperature $T_c$ and the
critical exponents for the disordered chain with $\sigma = 0.75$.  In
Fig.~\ref{binder}, we show rescaled data for the Binder ratio $g$ as a
function of temperature for different system sizes [see Eq.~(\ref{g})].  
The data scale well for $T_c = 0.62 \pm 0.03$ and $1/\nu = 0.30 \pm 0.03$,
and agree with previous results by Leuzzi.\cite{leuzzi:99}
The inset of the aforementioned figure shows the unscaled data for $g$
which cross at $T_c \approx 0.62$. The error bars are estimated by varying
$g$ slightly until the data do not collapse well.

\begin{figure}
\centerline{\epsfxsize=\columnwidth \epsfbox{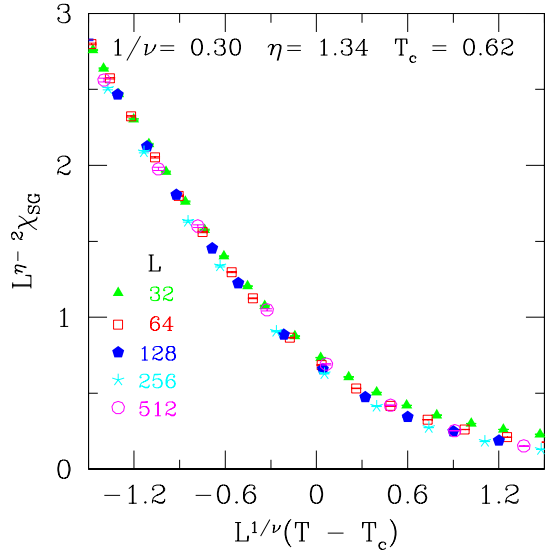}}
\vspace{-1.0cm}
\caption{
Data for the rescaled spin-glass susceptibility $L^{\eta - 2}\chisg$  
for different sizes as a function of $L^{1/\nu}(T - T_c)$. The data 
scale well for $1/\nu = 0.30 \pm 0.03$, $T_c = 0.62 \pm 0.03$, and
$\eta = 1.34 \pm 0.03$.
}
\label{chisg}
\end{figure}

We use the same approach below for the spin-glass susceptibility, for
which the data, rescaled according to Eq.~(\ref{chi}), are presented in
Fig.~\ref{chisg}. The data scale well for $\eta = 1.34 \pm 0.03$ and the
previously mentioned values of $1/\nu$ and $T_c$. These values can be
compared with the results of Kotliar {\em et al.}:\cite{kotliar:83} $\eta
= 3 - 2\sigma \ (= 1.5)$, and $1/\nu = (1/3) - 4 \epsilon$ to first order
in $\epsilon$, where $\epsilon = \sigma - 2/3$. The latter gives $1/\nu =
0$ suggesting that truncating the expansion to first order in $\epsilon$
is not valid for $\sigma = 3/4$.

Having studied the critical region of the model to estimate $T_c$ we are
now able to simulate the system at low temperatures ($T \approx 0.16T_c$)
in order to probe the spin-glass phase and test if $\theta = \theta^\prime$.

\label{results1}

\section{Results at Low Temperatures}
\label{results2}

In this section we study the model with $\sigma = 0.75, d=1$, marked by
a cross in Fig.~\ref{dsigma}, at temperatures well below $T_c \simeq
0.62$, in order to get information about large-scale, low-energy
excitations in the spin-glass state.

Differences between the various models for the spin-glass state can be
quantified by
studying\cite{marinari:00,katzgraber:01,reger:90,marinari:98a,marinari:99}
$P(q)$, the distribution of the spin-glass overlap $q$. The RSB picture predicts a
nontrivial distribution with a finite weight in the tail down to $q = 0$,
independent of system size. On the contrary, the droplet picture predicts
that $P(q)$ should be trivial with only two peaks at $\pm q_{\rm EA}$,
where $q_{\rm EA}$ is the Edwards-Anderson order parameter. For finite
systems there is also a tail down to $q = 0$ but this vanishes in the
thermodynamic limit like\cite{fisher:86,bray:86,moore:98} $\sim
L^{-\theta'}$ with $\theta' = \theta$.

\begin{figure}
\centerline{\epsfxsize=\columnwidth \epsfbox{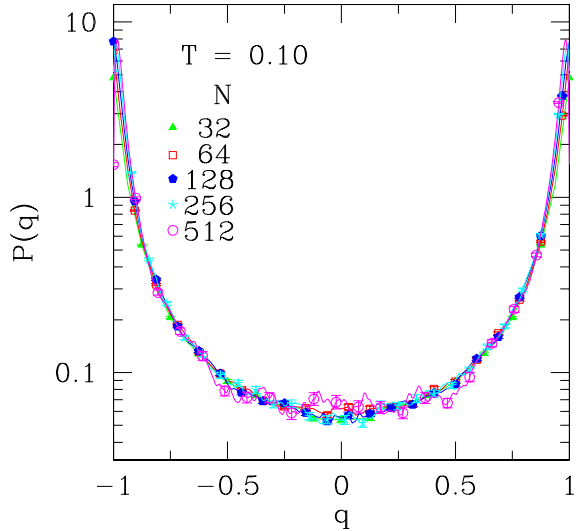}}
\vspace{-1.0cm}
\caption{
Data for the overlap distribution $P(q)$ at $T=0.10$. Note that the
vertical scale is logarithmic to better make visible both the peak at
large $q$ and the tail down to $q=0$. In this, as well as in
Fig.~\ref{pq0.23}, we only display {\em some}\/ of the data points as
symbols, for clarity, but the lines connect {\em all}\/ the data points.
This accounts for the curvature in some of the lines between neighboring
symbols. In this paper, all distributions are normalized so that the area
shown under the curve is unity.
}
\label{pq0.10}
\end{figure}

\begin{figure}
\centerline{\epsfxsize=\columnwidth \epsfbox{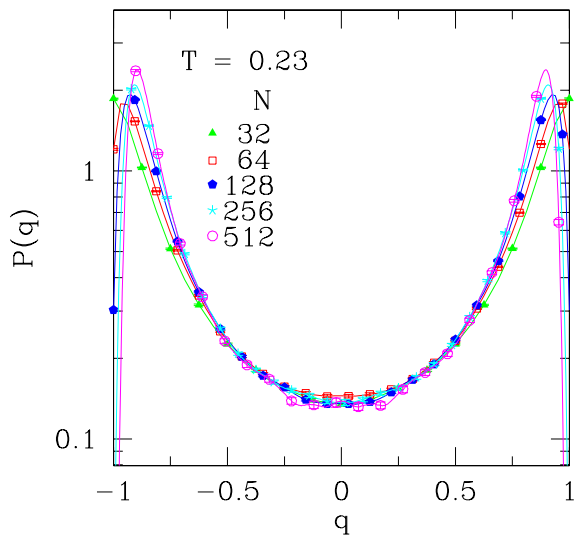}}
\vspace{-1.0cm}
\caption{
Same as for Fig.~\ref{pq0.10} but at $T=0.23$.
}
\label{pq0.23}
\end{figure}

\begin{figure}
\centerline{\epsfxsize=\columnwidth \epsfbox{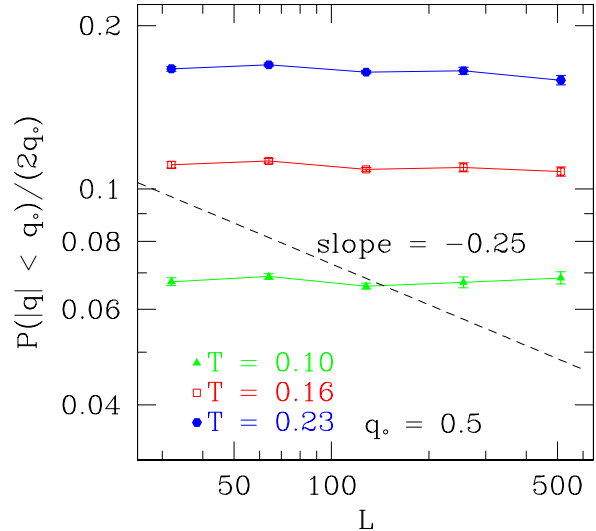}}
\vspace{-1.0cm}
\caption{
Log-log plot of $P(|q| < q_\circ)/(2q_\circ)$, the spin overlap
distribution for small $q$, against $L$ for $q_\circ = 0.50$. The data are
observed to not depend significantly on the system size $L$. The dashed
line has slope $-0.25$, the theoretical prediction of the droplet 
model (Ref.~\onlinecite{fisher:88}).
Asymptotically, the data should be parallel to this line according to the
droplet theory.
}
\label{pofnull}
\end{figure}

Figures \ref{pq0.10} and \ref{pq0.23} show data for $P(q)$ at temperatures
0.10 and 0.23, respectively. There is clearly a peak for large $q$ and a
tail down to $q=0$. At both temperatures one sees that the tail in the
distribution is essentially independent of system size. A more precise
determination of the size dependence of $P(0)$ is shown in
Fig.~\ref{pofnull} where, to improve statistics, we average over the
$q$ values with $|q| < q_\circ$, with $q_\circ=0.50$. The
expected\cite{fisher:86,bray:86,moore:98} behavior in the droplet model is
$P(0) \sim L^{-\theta'}$, with $\theta' = \theta$
where\cite{bray:86b,fisher:88} $\theta = d - \sigma$. The dashed line in
Fig.~\ref{pofnull} has slope $-0.25$, the expected value for $\sigma =
0.75$ according to the droplet model. The size dependence is consistent
with a constant $P(0)$, which implies that the energy to create a large
excitation does not increase with size, at least for the range of sizes
studied here.

We perform a two-parameter fit of the data in Fig.~\ref{pofnull} of the
form $a L^{-\theta'}$ to find for $T = 0.10$, $\theta' = -0.005 \pm 0.009$,
for $T = 0.16$, $\theta' = -0.020 \pm 0.007$, and for $T = 0.23$, $\theta' =
-0.016 \pm 0.006$. The fitting probabilities\cite{press:95} $Q$ for the
different temperatures are $0.135$, $0.186$, and $0.08$, respectively,
which is reasonable. Fixing $\theta' = 0$, we obtain fitting probabilities
of $0.214$, $0.053$, and $0.076$, respectively. On the contrary, fixing
$\theta'= 0.25$, the predicted value for the droplet picture,
we find $Q \lesssim 10^{-31}$ for
all three temperatures. Hence we are confident that, within error bars,
$\theta' \simeq 0$ for the range of sizes studied.

Moore \cite{moore:86} has argued that
$P(q)$ should be trivial in the thermodynamic limit
(i.e., just two delta functions at $q = \pm q_{EA}$) 
to first order in an expansion about the ``lower critical dimension''
$\sigma = 1$. This seems to disagree with our numerics, although possibly
even larger sizes might be necessary to see the
asymptotic behavior.

\section{Domain-wall energies at zero temperature}
\label{resultsT0}

Since there are analytic predictions, Eqs.~(\ref{theta_lr}) and
(\ref{theta_SR}), for $\theta$ for the present model, it is useful to
directly calculate $\theta$ from zero-temperature domain-wall calculations
for different values of $\sigma$.

We use parallel tempering here as an optimization algorithm. We find that
if we take the lowest temperature $T_{\rm min}$ to be $0.05$, then the
minimum-energy state found at this temperature is the ground state. To
test whether the true ground state has been reached, two criteria were
adopted: (i) the same minimum-energy state has to be reached from two
replicas at $T_{\rm min}$ for all samples, and (ii) this state has to be
reached during the first $1/4$ of the sweeps in both copies.  These
conditions were satisfied for the parameters used in the simulations which
are shown in Table \ref{simparamst0}.

\begin{table}
\caption{
Parameters of the $T = 0$ simulations. The table shows the total number of
Monte Carlo steps used for each value of $\sigma$ and $L$.
\label{simparamst0}
}
\begin{tabular*}{\columnwidth}{@{\extracolsep{\fill}} c c c c c c }
\hline
\hline
$\sigma$ &  $L = 16$  & $L = 32$ & $L = 64$ & $L = 128$ & $L = 256$ \\
\hline
0.10 & $2\times 10^3$ & $4\times 10^3$ & $8\times 10^3$ & $4\times 10^4$ & $12\times 10^4$ \\
0.25 & $2\times 10^3$ & $4\times 10^3$ & $8\times 10^3$ & $4\times 10^4$ & $12\times 10^4$ \\
0.50 & $2\times 10^3$ & $4\times 10^3$ & $8\times 10^3$ & $4\times 10^4$ & $12\times 10^4$ \\
0.62 & $2\times 10^3$ & $4\times 10^3$ & $8\times 10^3$ & $4\times 10^4$ & $12\times 10^4$ \\
0.75 & $2\times 10^3$ & $4\times 10^3$ & $8\times 10^3$ & $4\times 10^4$ & $12\times 10^4$ \\
0.87 & $2\times 10^3$ & $4\times 10^3$ & $8\times 10^3$ & $4\times 10^4$ & $12\times 10^4$ \\
1.00 & $2\times 10^3$ & $4\times 10^3$ & $8\times 10^3$ & $8\times 10^4$ & $6\times 10^5$ \\
1.25 & $2\times 10^3$ & $4\times 10^3$ & $6\times 10^4$ & $6\times 10^5$ & \\
1.50 & $2\times 10^3$ & $4\times 10^3$ & $6\times 10^4$ & & \\
1.75 & $2\times 10^3$ & $4\times 10^3$ & $6\times 10^4$ & & \\
2.00 & $2\times 10^3$ & $4\times 10^3$ & $6\times 10^4$ & & \\
2.50 & $2\times 10^3$ & $4\times 10^3$ & $2\times 10^5$ & & \\
3.00 & $2\times 10^3$ & $4\times 10^3$ & $2\times 10^5$ & & \\
\hline
\hline
\end{tabular*}
\end{table}

Having found the ground state with periodic (P) boundary conditions, we
then flip the boundary conditions to ``antiperiodic'' (AP) by the
following prescription. Consider one of the nearest neighbor (nn) bonds,
e.g., that between sites 1 and $L$ in Fig.~\ref{chain}. Then, for all $i$
and $j$, change the sign of the bond between $i$ and $j$ if the shorter
path between those sites goes through the chosen nn bond. The new ground
state is calculated and note is taken of the difference in energy
\begin{equation}
\delta E = E_{\rm AP} - E_{\rm P}
\end{equation}
for each sample. On average, there is no preference for AP or P and so
$[\delta E]_{\rm av} = 0$. Hence we take the average of the absolute value
\begin{equation}
\Delta E = [\,|\delta E|\,]_{\rm av}
\end{equation}
as a measure of the characteristic domain-wall energy.

\begin{figure}
\centerline{\epsfxsize=\columnwidth \epsfbox{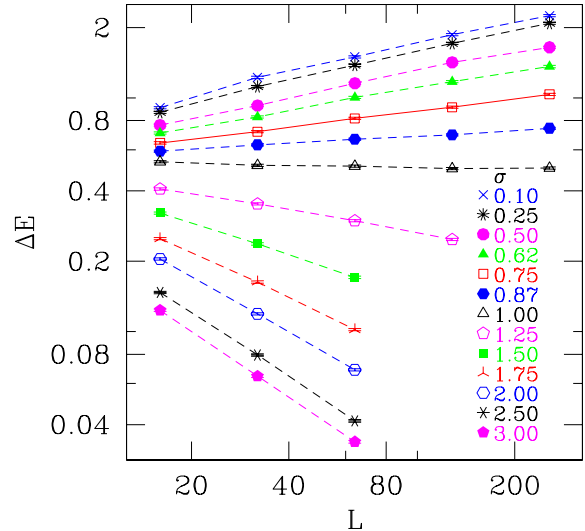}}
\vspace{-1.0cm}
\caption{
Data for the defect energy $\Delta E = [\,|E_{\rm P} - E_{\rm AP}|\,]_{\rm
av}$ for different sizes and values of $\sigma$. The solid lines connect
the data for $\sigma = 0.75$, the value of $\sigma$ studied at finite
temperatures.
}
\label{de}
\end{figure}

Data for a range of sizes and values of $\sigma$ are shown in
Fig.~\ref{de}. Our program automatically scales the interactions to set
$T_c^{\rm MF}=1$, so no modifications are needed for the case $\sigma \le 0.5$
since the fact that the scale factor $c(\sigma)$ [in Eqs.~(\ref{bonds}]
and (\ref{tcmf})) tends to zero for $L \to \infty$ in this range, is taken
care of automatically.

A smaller range of sizes has been studied for larger $\sigma$ as parallel
tempering is less efficient in finding the ground state in this limit. At
first sight this seems surprising because the nearest-neighbor
interactions dominate and so there is no frustration, apart possibly from
that due to the boundary conditions. The spins become stuck when a
frustrated nn bond is larger in magnitude than the neighboring bonds.  
Parallel tempering does not seem to help much in this situation, perhaps
because the local minima do not have a hierarchical structure. It would be
interesting to investigate in more detail the conditions under which
parallel tempering is efficient.

\begin{figure}
\centerline{\epsfxsize=\columnwidth \epsfbox{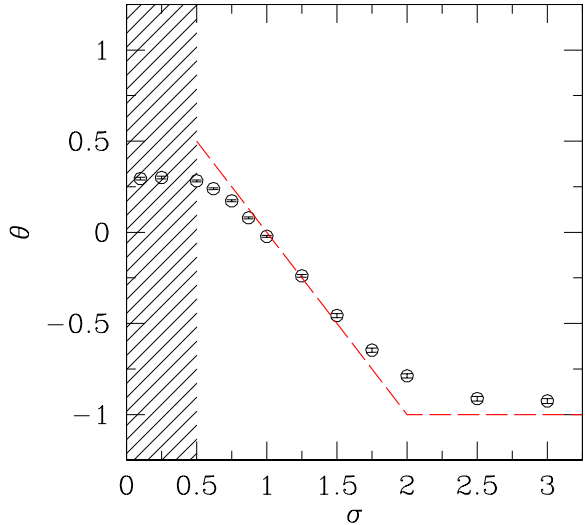}}
\vspace{-1.0cm}
\caption{
Variation of $\theta$, with $\sigma$, obtained from fitting the data in
Fig.~\ref{de} to $\Delta E \sim L^\theta$. Also shown is the analytic
prediction (dashed line): $\theta = 1 - \sigma$ for $1/2 < \sigma < 2$,
and $\theta = -1$ for $\sigma > 2$. The shaded region is where there is no
thermodynamic limit unless the interactions are scaled appropriately by
system size.
}
\label{theta}
\end{figure}

One expects that $\Delta E \sim L^\theta$, where $\theta = 1 - \sigma$ in
the long-range region $1/2 < \sigma < 2$, and $\theta = -1$ in the
short-range region $\sigma > 2$. Fits to the data obtained in
Fig.~\ref{de}, along with the analytic prediction, are shown in
Fig.~\ref{theta} and in Table \ref{theta-fits}. In particular, there is a
distinctive positive slope for $\sigma = 0.75$ in Fig.~\ref{de} (solid
line), corresponding to a positive value of $\theta$. This is in clear
contrast to the finite-$T$ data shown in Fig.~\ref{pofnull}.

\begin{table}
\caption{
Fits for $\Delta E \sim aL^{\theta}$ for different values of $\sigma$. $Q$
represents the quality of fit. The last column in the table shows the
range of values of $L$ used in the fits (for the possible values see Table
\ref{simparamst0}). For $0.10 \le \sigma \le 0.25$, we exclude the data for
$L = 16$ from the fits due to finite-size effects for small values of
$\sigma$.
\label{theta-fits}
}
\begin{tabular*}{\columnwidth}{@{\extracolsep{\fill}} c r r r c }
\hline
\hline
$\sigma$ &  $a$  & $\theta$ & $Q$ & $L$ Range\\
\hline
0.10 & $-0.818\pm 0.033$ & $ 0.295 \pm 0.007$ & $0.495$ & 32 -- 256\\
0.25 & $-0.928\pm 0.032$ & $ 0.301 \pm 0.007$ & $0.759$ & 32 -- 256\\
0.50 & $-1.023\pm 0.033$ & $ 0.282 \pm 0.005$ & $0.007$ & 16 -- 256\\
0.62 & $-1.007\pm 0.022$ & $ 0.239 \pm 0.005$ & $0.235$ & 16 -- 256\\
0.75 & $-0.925\pm 0.022$ & $ 0.173 \pm 0.005$ & $0.852$ & 16 -- 256\\
0.87 & $-0.741\pm 0.021$ & $ 0.079 \pm 0.005$ & $0.801$ & 16 -- 256\\
1.00 & $-0.577\pm 0.022$ & $-0.023 \pm 0.005$ & $0.319$ & 16 -- 256\\
1.25 & $-0.225\pm 0.028$ & $-0.239 \pm 0.007$ & $0.222$ & 16 -- 128\\
1.50 & $ 0.137\pm 0.030$ & $-0.457 \pm 0.011$ & $0.349$ & 16 -- 64\\
1.75 & $ 0.414\pm 0.041$ & $-0.647 \pm 0.012$ & $0.150$ & 16 -- 64\\
2.00 & $ 0.600\pm 0.042$ & $-0.788 \pm 0.012$ & $0.727$ & 16 -- 64\\
2.50 & $ 0.619\pm 0.044$ & $-0.913 \pm 0.013$ & $0.289$ & 16 -- 64\\
3.00 & $ 0.499\pm 0.045$ & $-0.935 \pm 0.013$ & $0.975$ & 16 -- 64\\
\hline
\hline
\end{tabular*}
\end{table}

We see that the overall trends in the analytic prediction are reproduced
by the data, but there is some rounding where the analytical result has a
change in slope. Note that our results for $\theta$ always lie well below
$1/2$, the predicted\cite{bray:86b,fisher:88} value for $\sigma = 1/2$,
and appear to saturate at about $0.3$ as $\sigma \to 0$, the SK limit.

\begin{figure}
\centerline{\epsfxsize=\columnwidth \epsfbox{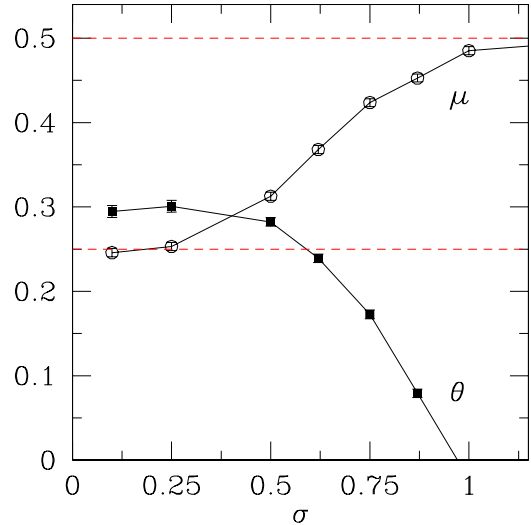}}
\vspace{-1.0cm}
\caption{
Defect energy exponent $\theta$ and the exponent for the rms
sample-to-sample fluctuations in the ground-state energy, $\mu$, for
different values of $\sigma$.
}
\label{de_def}
\end{figure}

In recent work Aspelmeier {\em et al}.\cite{aspelmeier:02} have given analytic
arguments, based on the RSB theory, that the
defect energy should vary as $N^\mu$ for short-range models in high
dimension $d$,
where $N=L^d$ and $\mu$ is the exponent for the rms sample-to-sample
fluctuations of the free energy of the SK model. A number of
authors\cite{cabasino:88,bouchaud:02,palassini:03} have found that $\mu
\simeq 1/4$, at least at $T=0$.
We have therefore calculated $\mu$ for our one-dimensional
model at $T=0$ for different values of $\sigma$
and show the results in Fig.~\ref{de_def}. Indeed
we find that $\mu \simeq 1/4$ for $\sigma \to 0$ (the SK limit). Also
note
that it crosses over to the expected ``self-averaging'' value of $\mu
=1/2$ in the region $\sigma > 1/2$ where we no longer need to scale the
interactions with system size. Since we work in $d=1$ the expectation from
Aspelmeier {\em et al}.\cite{aspelmeier:02} is that $\theta \to \mu$ for
$\sigma \to 0$. 
Figure \ref{de_def} also shows $\theta$ and, indeed,
it is quite striking how $\theta$ becomes close to $\mu$ at small
$\sigma$.

\section{Conclusions}
\label{conclusions}

We have performed finite-$T$ Monte Carlo simulations and ground-state
calculations of the long-range disordered Ising chain in one dimension
with power-law interactions. For the ground-state calculations we have
studied a range of values of $\sigma$ that encompasses all the possible
regimes, from large $\sigma$ where the system is controlled by the
short-range part of the interactions and has $T_c = 0$, all the way to
near the SK model limit, $\sigma = 0$. For the finite-temperature
Monte Carlo runs we have
concentrated on a single value of $\sigma$, namely $0.75$, which is in the
range where there is a finite $T_c$.

For $\sigma = 0.75$ we find the stiffness exponent for droplets,
$\theta'$, inferred from Monte Carlo simulations for sizes up to $L=512$,
to be close to zero,
whereas the stiffness exponent for domain walls, $\theta$, is positive.
We find $\theta = 0.173 \pm 0.005 $ (only statistical errors included)
somewhat lower than the analytic prediction\cite{bray:86b,fisher:88} that
$\theta = 0.25$.
These results are consistent with the RSB and TNT scenarios according to which
$\theta'=0$ and $\theta \ne \theta'$. Similar results have 
been found in earlier work on short-range models in three and four
dimensions.\cite{krzakala:00,palassini:00,katzgraber:01} According to the
droplet picture, $\theta' = \theta$ asymptotically, 
and there has
been discussion as to whether the failure of the numerics to see this
is an artifact of the small lattice
sizes,\cite{moore:02} or whether it will persist in the thermodynamic
limit. In the present work, the disagreement persists even though we are able
to study a much
larger range of $L$.

Also noteworthy is our finding that $\theta$ tends to a value of about 0.3
for $\sigma \to 0$, the SK model limit. This is below the droplet theory
prediction that $\theta$ should tend to 1/2 for $\sigma \to 0$, but is close
to the prediction of Aspelmeier {\em et al}.\cite{aspelmeier:02} 
that $\theta \to 1/4$ in this limit.

\begin{acknowledgments}
We would like to that M.~A.~Moore for stimulating discussions and for
bringing Ref.~\onlinecite{moore:86}
to our attention, and G.~Blatter for carefully reading the manuscript. 
H.G.K.~acknowledges support from the National Science
Foundation under Grant No.~DMR 9985978 and the Swiss National Science
Foundation.  A.P.Y.~acknowledges support from the National Science Foundation
under Grant No.~DMR 0086287, and the EPSRC under Grant GR/R37869/01. He
would also like to thank David Sherrington for hospitality during his stay
at Oxford. This research was supported in part by NSF cooperative
agreement ACI-9619020 through computing resources provided by the National
Partnership for Advanced Computational Infrastructure at the San Diego
Supercomputer Center. We would like to thank the University of New Mexico
for access to their Albuquerque High Performance Computing Center.  This
work utilized the UNM-Alliance Los Lobos Supercluster. Part of the
simulations were performed on the Asgard cluster at ETH Z\"urich.

\end{acknowledgments}

\bibliography{refs,comments}

\end{document}